\documentclass[twocolumn,amsmath,amssymb,amsthm,showpacs]{revtex4}
\usepackage[latin1]{inputenc}
\usepackage{mathrsfs}
\usepackage{graphicx,epsf,subfigure,pstricks}
\usepackage{color}

\begin{document}

\author{Jo\"el Marthelot$^{1,2}$}
\email{Joel.Marthelot@espci.fr}
\author{Beno\^it Roman$^1$}
\author{Jos\'e Bico$^1$}
\author{J\'er\'emie Teisseire$^2$}
\author{Davy Dalmas$^2$}
\author{Francisco Melo$^3$}
\title{Self-replicating cracks: a collaborative fracture mode in thin films}

\affiliation{$^1$PMMH, CNRS UMR 7636, UPMC Universit\'e Paris 6 \& Universit\'e Paris Diderot Paris 7, ESPCI-ParisTech, 10 rue Vauquelin, 75231 Paris Cedex 05, France.\\
$^2$SVI, CNRS UMR 125, Saint-Gobain Recherche, BP 135, 93303 Aubervilliers Cedex, France.\\
$^3$Departamento de F\'isica Universidad de Santiago de Chile, Avenida Ecuador 3493, 9170124 Estaci\'on Central, Santiago, Chile.
}
\date{\today}
\begin{abstract}
Straight cracks are observed in thin coatings under residual tensile stress, resulting into the classical network pattern observed in china crockery, old paintings or dry mud. Here, we present a novel fracture mechanism where delamination and propagation occur simultaneously, leading to the spontaneous self-replication of an initial template. 
Surprisingly, this mechanism is active below the standard critical tensile load for channel cracks and selects a robust interaction length scale on the order of 30 times the film thickness. Depending on triggering mechanisms, crescent alleys, spirals or long bands are generated over a wide range of experimental parameters. We describe with a simple physical model the selection of the fracture path and provide a configuration diagram displaying the different failure modes.
\end{abstract}

\pacs{62.20.mm, 68.60.Bs, 81.16.Rf, 81.20.Fw}

\maketitle 

Nanometer to micrometer thin film coatings are extensively used in material science to protect and functionalize surfaces~\cite{gioia97},  
from traditional thermal  barriers~\cite{padture02}, mechanical or chemical protection, to more recent applications in biomedical ~\cite{Haldar2006}  
or stretchable electronics~\cite{rogers10}. However, deposition processes, thermal expansion mismatch or simply mechanical loading generally result into compressive or tensile residual stresses that induce two main types of failure of coatings.
Compressive stresses commonly induce the formation of wrinkles~\cite{bowden98} and blisters~\cite{gioia97, vella09}
whereas tensile stresses leads to straight {\it channel cracks} across the film thickness~\cite{Hutchinson92}. 
Once triggered, these fractures propagate along a straight trajectory,
being only deflected in the close vicinity of a previous fracture path, where they tend to connect perpendicularly to the free boundary. 
In the case of a stiff substrate this interaction distance is on the order of the thickness of the coating. 
Such familiar hierarchical disordered patterns are for instance observed in dry mud~\cite{Atkinson91, Shorlin00}, in the glaze of ceramics or even in plant venation and urban networks~\cite{Bohn05}. 
Delamination may eventually occur after the previous fracture pattern has been established~\cite{Lazarus11}.  
\begin{figure}[!h]
\begin{center}
\includegraphics[width=\columnwidth]{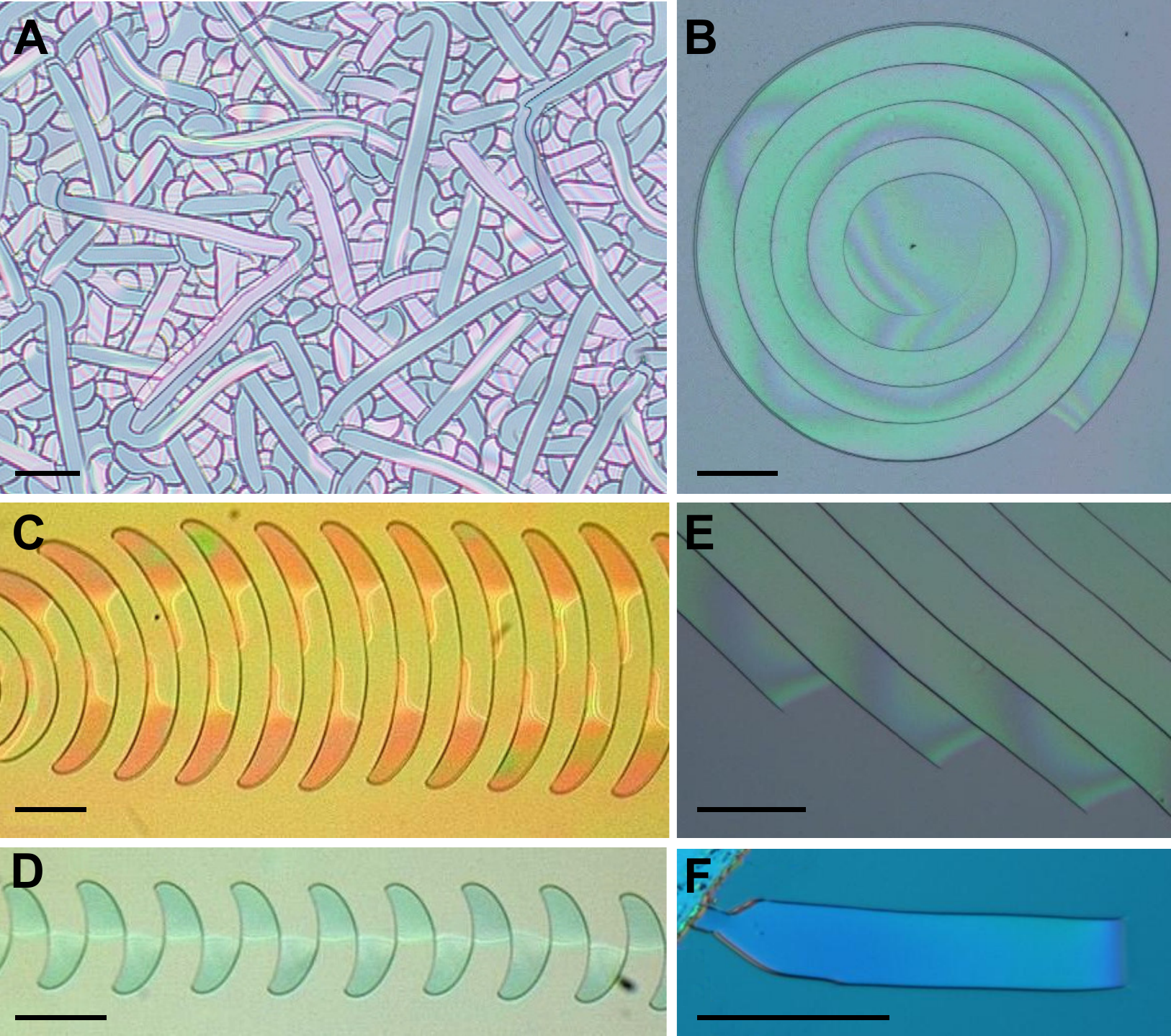}
\vspace{-5mm}
\caption{Unusual cracks in thin film moderately adherent to a substrate (scale bar $100 \mu$m). Numerous nucleation spots lead to complex patterns (a). Self-replicating cracks triggered by scarce defects (see the supporting movies S1, S2, S3 in the Supplemental Material): Archimedean spirals (b); regular alleys of crescents (c),(d); and parallel bands (e) follow an initial arch, loop, or line, at a fixed distance W 1 . (f) Pairs of cracks simultaneously follow each other path leading to isolated bands.}
\end{center}
\end{figure}

This scenario is observed in our experiments with thick coatings of commercial {\it Spin-On-Glass} (SOG) trimethylsiloxane  when adhesion is enhanced by a
plasma treatment of the substrate.
In such coatings, which are commonly used {for adjusting the optical index of buffer layers in laser cavities}, residual bi-axial stresses result from a sol-gel process~\cite{Lebental07}.
 However, unexpected crack morphologies  (Fig.~1A) are observed in the case of moderate adhesion ($\Gamma\sim 0.5\,\mathrm{J/m}^2$). Decreasing the density of defects ({\it e.g.} by filtering the SOG solution prior to reaction) leads to intriguing patterns. Archimedean spirals (Fig.~1B),  crescent alleys (Fig.~1C-D) or parallel bands (Fig.~1E-F) grow spontaneously after the sol-gel condensation reaction on a silicon wafer as the sample is removed from the reacting chamber (see movies S1, S2, S3 in {\it Supplemental material}).
{These cracks are triggered by sporadic defects in the coating but can also be induced locally by the operator ({\it e.g.} several fractures randomly initiate from the scratch of a sharp blade).}
As they propagate, cracks tend to follow  a previous cut at a fixed distance.
Crescent alleys, spirals and series of parallel bands thus correspond to the self-replication of an initial arch, loop or line, respectively. 
The growth of an isolated band can also be viewed as a pair of cracks following simultaneously each other's path.  
 
These patterns are strikingly different from usual crazing glaze figures,
but they are 
not specific to the SOG system studied in this article. 
Indeed, similar crescent alleys or spiral patterns have been mentioned  in different areas of material science~\cite{Lebental07,Sendova03, Wan09, Wu13, Bozzini12,Meyer04,bursikova06,Malzbender00}. 
However, the corresponding fracture mechanism remains mysterious.\\

We carried experiments on SOG layers with different thicknesses and adhesion properties (deposition and characterization of the coating are decribed in {\it Supplemental Material}).
We present in Fig.~2 the characteristic length scale of the patterns observed with SOG on silicon. 
More precisely, this length scale corresponds to the crack replicating distance $W_1$ in the case of spirals and crescent alleys, and to the width $W_2$ of isolated bands (see inset images in Fig.~2). 
We extend these data with measurements extracted from the literature, and with additional experiments conducted with macroscopic layers of varnish.
As a striking result, the scale of the patterns is proportional to the thickness of the film $h$  over 4 orders of magnitude. 
We indeed obtain $W_1 \simeq 32 h$  and $W_2 \simeq 25 h$, which
is large in comparison with the interaction length of the usual channel cracks (on the order of $h$). 
 The robustness of these patterns observed with very different types of coatings and deposition methods,  such as sol-gel~\cite{Lebental07, Wu13,Sendova03, Wan09}, magneton sputtering~\cite{Meyer04}, or evaporation~\cite{Bozzini12} suggests that their characteristic width is independent from both loading conditions or material properties and only depends on the thickness of the film. 
This robust size selection clearly indicates that these patterns are 
 different from other spiral or oscillating fracture paths observed in systems involving thermal gradients~\cite{Yuse93}, drying fronts~\cite{Leung01} or tearing with a blunt object~\cite{Audoly05}.

\begin{figure}[htb]
\centering
  \includegraphics[width=\columnwidth]{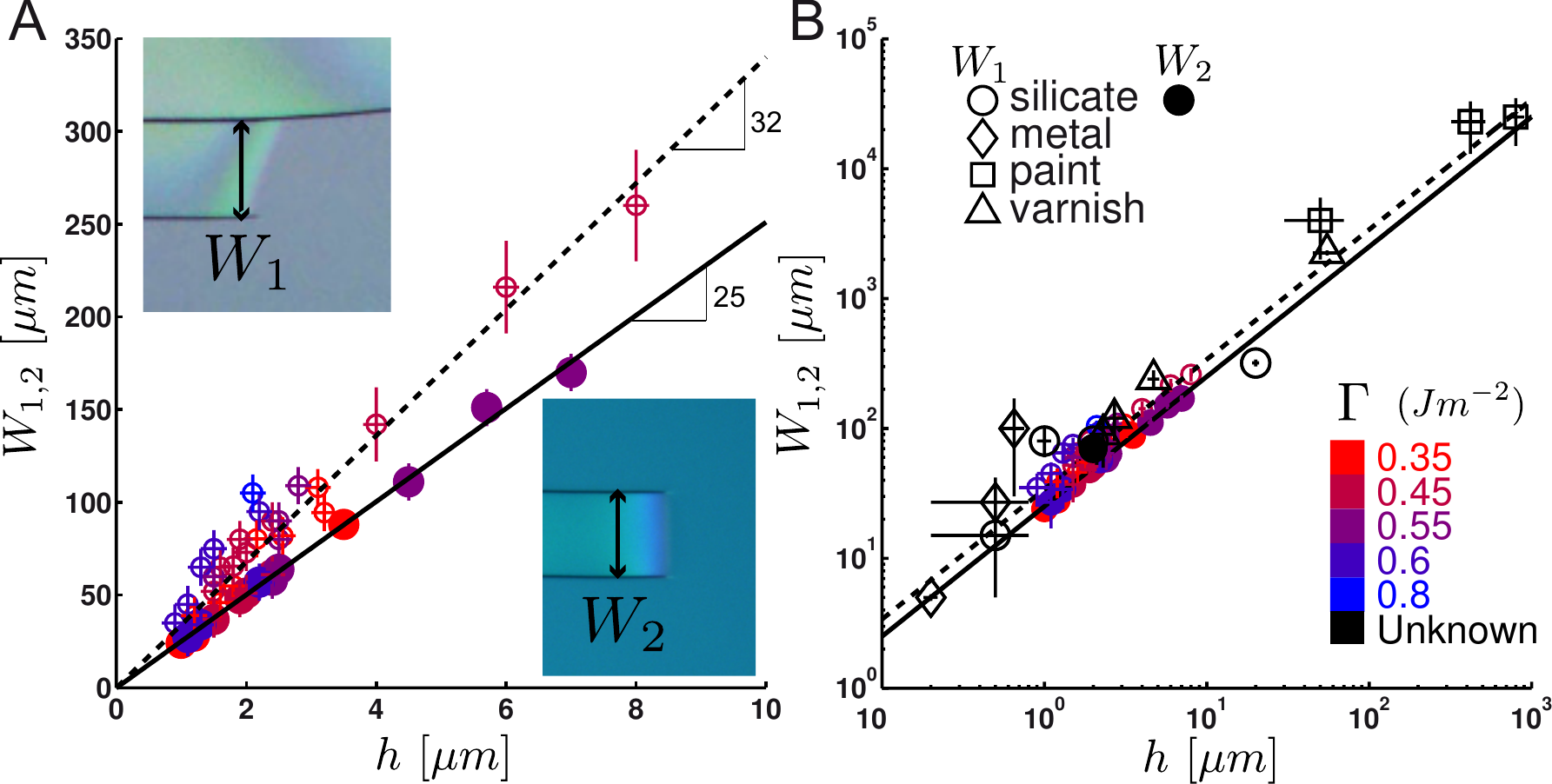}
  \vspace{-5mm}
   \caption{({\bf A}) Width of the delaminating front  in  SOG films with different adhesion energies $\Gamma_0$ (color coded) as a function of the thickness $h$ of the film for different morphologies: pitch of the spiral or wavelength of the crescent alleys  ($W_1$, open symbols) and width of paired cracks ($W_2$, filled symbols). Lines correspond to linear fits of the data. ({\bf B}) Generalization to a wider range of systems including data from the literature (other silicate films \cite{Sendova03, Wan09, Malzbender00, Lebental07}, metal films \cite{Meyer04, Bozzini12, bursikova06}) and macroscopic measurements on varnish or paint.
 }
\end{figure}

Our experimental system also allows for live observation of the quasistatic crack dynamics using a reflection microscope.  Interferences fringes show that the film simultaneously delaminates as the crack propagates. {FEG-SEM imaging shows that the fracture fully extends to the bottom of the debonding layer (see {\it Supplemental material}).} The film may eventually re-adhere to the substrate far from the crack front. Delamination is thus not always obvious in {\it post-mortem} images of the cracked coating, although it actually takes place during propagation {(as confirmed by AFM images, see {\it Supplemental material}).}\\

Before describing these peculiar cracks, we first recall the case of classical channel cracks propagating through a thin film under an isotropic tensile stress $\sigma$. Following Griffith classical criterion, fracture propagates if the elastic energy released per unit length overcomes the fracture energy,  \begin{equation}
2 \gamma h e \geq G_c h,~~\mathrm{with}~~e= h\sigma^2(1-\nu)/E,
\label{eq:channel}
\end{equation}
where  $e$ is the elastic energy in the film per unit surface ($h, E$ and $\nu$ are respectively the thickness, Young modulus and Poisson ratio of the film) and $G_c$ is the fracture energy per unit area.
The coefficient $\gamma$ depends on the mismatch in elastic properties  between the film and the substrate and is of order 1 for the relevant case of comparable rigidities~\cite{Hutchinson92}. 
In physical terms, $\gamma h$  gives the lateral size over which fracture allows stresses to relax in the bonded coating.
 Eq.~\ref{eq:channel} shows that classical channel cracks are expected to propagate above a critical thickness $h_c = G_c E/2\gamma\sigma^2(1-\nu)$.
Surprisingly, the non-standard crack patterns displayed in Fig.~1 are observed below $h_c$, confirming a different fracture mechanism. In fact fracture and delamination collaborate here, releasing residual stresses in the large delaminated  area surrounded by free boundaries. 

We now present a simple theoretical framework for this new collaborative mode, which explains the robustness of the fracture path geometries, and provide a general diagram for its domain of existence. 
We first  focus on the simpler case of an isolated band, {\it i.e.} a pair of cracks propagating simultaneously (Fig.~1F), and then describe path-following cracks (Fig.~1B-E).

We consider a pair of cracks propagating in a local direction $\theta$ along a symmetric, but arbitrary, path with curvilinear length $s$ (Fig.~3A).
In addition to the energy released along the edges $2\gamma eh s$, which would drive the propagation of isolated channel cracks (Eq.~\ref{eq:channel}),  debonding an area $A$ is expected to completely release the residual energy $eA$, except in the vicinity of the debonding front where  boundary conditions maintain strains parallel to the front.
This incomplete release extends in an area proportional to $W^2$ leaving a residual energy $\alpha e W^2$ (Fig.~3A).
Nevertheless, the delaminated film is also free to tilt up along the debonding front, releasing stresses
perpendicular to the delamination front (Fig.~3B). This effects extends on a distance of order~$h$ ahead of the front, corresponding  an additional energy release $\beta ehW$.  
Summing the different terms finally gives the released energy
 \begin{equation}
 \mathcal{E}_r = e(A - \alpha  W^2 +\beta hW + 2 \gamma h s), \label{eq:Er}
 \end{equation}
 where $\alpha,\beta,\gamma$ are non-dimensional constants.
This expression is  in agreement with experimental measurement of the strain fields obtained from image correlation and with finite element calculations conducted on a parallel band cut from a pre-strained film adhering to a substrate (see {\it Supplemental material}). 

 \begin{figure}[h]
\centering
  \includegraphics[width=\columnwidth]{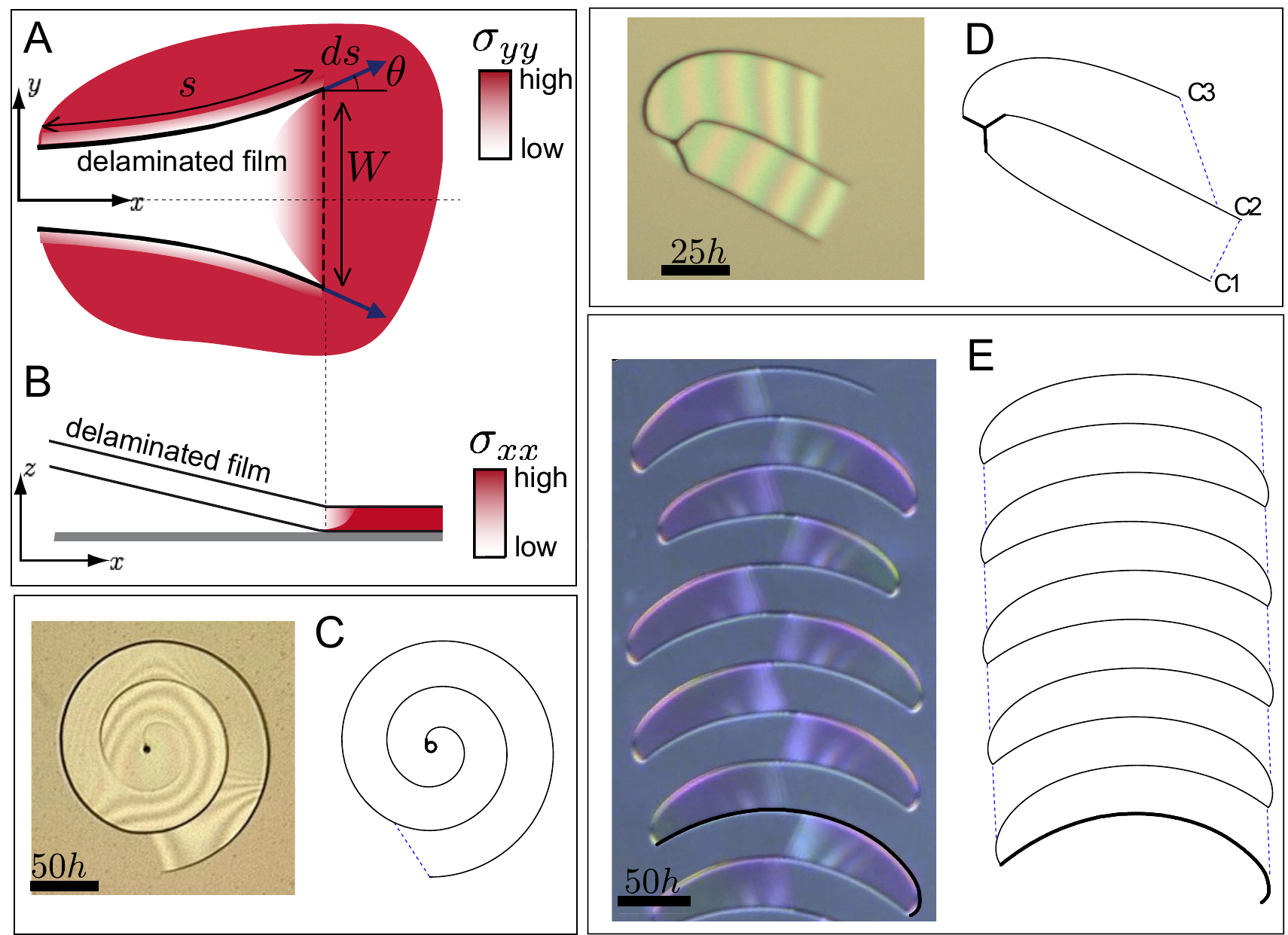}
  \caption{ 
({\bf A}) Delaminated symmetric band with imperfect release of  stress $\sigma_{yy}$ close to the debonding front   ({\bf B}) The spontaneous tilt of the band releases stresses $\sigma_{xx}$ ahead of the debonded zone. ({\bf C-E}) Comparison of experimental crack paths with the prediction from energy minimization  with ($\alpha=0.0251,\beta=1.26$) starting from an initial crack represented by a thick black line. 
The debonding front is drawn in dashed blue line.
  }
  \label{fig-energy}
\end{figure}

Following Griffith's criterion, symmetric cracks should propagate by $ds$ along a given direction $\theta$ when adhesion and fracture energies, $(\Gamma W\cos\theta + 2 G_c h)ds$, are balanced by  the released elastic energy $d\mathcal{E}_r$, {\it i.e.} for:
\begin{equation}
e [ W\!\cos\theta + 2 \gamma  h -2(2\alpha W - \beta h)\!\sin\theta ] =\Gamma W\!\cos\theta + 2G_c h,
\label{eq:griffith}
\end{equation}
where we have used the geometrical relations $dA=W ds\cos\theta$ and $dW=2ds\sin\theta$.\\
In addition,  
cracks are expected to propagate in the direction that maximizes the energy release rate~\cite{Hutchinson92,Cotterell80},\begin{equation}
e W\sin\theta + 2e(2\alpha W -\beta h)\cos\theta = \Gamma W \sin\theta.
\label{eq:release}
\end{equation}
Combining equations~(\ref{eq:griffith}) and (\ref{eq:release})  finally leads to 
\begin{equation}
\sin\theta = -\frac{e}{G_c-\gamma e }(2\alpha W/h - \beta), \label{eq:sintheta}
\end{equation}
which predicts that the propagation is straight ($\theta=0$) for a specific inter-crack distance 
\begin{equation}
W_{2} = \frac{\beta}{2\alpha}h.  \label{eq:W2}
\end{equation}
The sign of $\theta$ in Eq.~\ref{eq:sintheta} tends to compensate any deviation from $W_{2}$ and leads to a stable mode of propagation for self peeling bands with constant width. 
In physical terms, this width optimizes the energy released (Eq.~\ref{eq:Er}), by balancing the surface energy  ($\alpha$ term) which penalizes large $W$, with the line energy ($\beta$ term) dominant at small $W$. Because the optimal width is set by elasticity alone, it is independent from the magnitude of loading,  adhesion or fracture energies, as observed in experiments (Fig.~2).
A numerical estimate of the parameters $\alpha$ and $\beta$, and the study of stress intensity factors, 
provide a predicted width $W_2=23.7h$ in quantitative agreement with experiments (see {\it Supplemental Material}).

Although the geometry of the bands is similar for all systems, the condition for propagation does depend on the magnitude of adhesion.
Indeed, the propagation of a strip is energetically possible (Eq.~\ref{eq:griffith}) if the adhesion energy is exactly $\Gamma  = \Gamma_2,$ with
\begin{equation}
\frac{\Gamma_2}{e}=  1 - 2\left(\frac{G_c}{e} -\gamma \right)\frac{h}{W_2}.
\label{eq:speed}
\end{equation}
Experimental measurements of the adhesion energy show that $\Gamma$ depends on the speed of propagation (see  {\it Supplemental material}). 
Starting from a minimum equilibrium value $\Gamma_0$, the adhesion energy $\Gamma$ is found to increase with the velocity of the front, as observed in similar systems where water activated delamination is limited by diffusion kinetics~\cite{Lin07}. As a consequence, the propagation of a pair of parallel cracks is only possible if $\Gamma_{2} > \Gamma_0$, and  $\Gamma(v)=\Gamma_2$ 
prescribes the propagation speed $v$, which typically ranges from 1 to 50~$\mu$m/s in our experiments. {Note that although $G_c$ is also expected to depend on $v$, we did not include this effect which only changes the numerical value of the selected velocity.}\\

We extend previous arguments to the case of crescent alleys or spirals, where a single crack follows an older fracture  path of arbitrary geometry, assuming the same expression for the energy released.
We derive the inclination of the debonding front (assumed straight), and obtain general analytical equations for the  crack trajectory and the conditions for propagation (see {\it Supplemental Material}). 
The resolution of these equations provide an excellent prediction of the experimental path (see Fig~3C,D,E).
Some particular analytical solutions are worth mentioning.
In the case of a crack interacting with a previous straight cut, we obtain a stable width,
\begin{equation}
W_1 = \frac{\beta h}{2\alpha}\left(\frac{1}{\sqrt{2}}+\frac{G_c/e  - \gamma} {2\beta} \right), \label{eq:W1}
\end{equation}
corresponding to a debonding front tilted by an angle of $45^{\circ}$.
This relation explains the robust tendency to replicate a crack path at a well defined distance observed in experiments (Fig.~3C),
although the quantitative dependence of $W_1$ with $G_c/e$ is difficult to capture (Fig.~2A).
The propagation velocity is now set by $\Gamma(v)=\Gamma_1$, with
\begin{equation}
 \frac{\Gamma_1}{e} =  1 -  \left(\frac{G_c}{e}-\gamma \right)\frac{h}{W_1}. \label{eq:speedW1}
\end{equation}
In the case of a crack rotating around a point (or around the tip of a straight segment), we obtain a circle of radius equal to the width $W_2$ of the symmetric band. Crack velocities are also identical, given by $\Gamma(v)=\Gamma_2$.  
Since in our system $\Gamma_2 < \Gamma_1$, cracks rotating around a tip, or paired parallel cracks, propagate at lower velocities than cracks following a straight cut. A crack following a parallel band thus catches up with the paired cracks as illustrated in Fig.~3D (see also movie S4 in {\it Supplemental material}).
 More importantly, in the case  where adhesion energy is high enough ($\Gamma_0>\Gamma_2$), rotation around a sharp turning point (such as the extreme point of the crescent path) is not energetically possible, and the front stops.
In experiments, we observe that a secondary delamination front then slowly develops, and triggers the propagation of the returning branch, leading to a crescent alley (Fig.~3E and movie~S2).\\

Three main physical ingredients dictate the different crack patterns : the residual energy density per unit surface $e$, the fracture energy of the film $G_c$ and the adhesion energy for vanishing speed $\Gamma_0$. 
In our experiments with silicate coatings, $e$ and $\Gamma_0$ could be varied independently by respectively adjusting the thickness of the film and the chemical treatment of the substrate, while $G_c$ was set by the system.
In Fig.~4 we present the morphologies of the cracks structures as a function of the two non-dimensional parameters, $\gamma e/G_c$ and $e/\Gamma_0$. 

In the classical picture, isolated channel cracks propagate when $\gamma e/G_c \geq 1/2$.
Delamination is energetically favorable when the residual elastic energy overcomes adhesion energy,
$
e\geq\Gamma_0,
$
but this mode of failure requires free boundaries to propagate and the damaged zone is usually confined within limited regions along defects or between channel cracks. 
The boundaries corresponding to these classical conditions are drawn as straight lines in Fig.~4, and the pink colored domain is therefore expected to be stable. 
However, the collaborative mechanism occurs within this region usually recognized as safe. 
Two additional limiting boundaries are introduced to describe these modes : $\Gamma_0 =\Gamma_2$ which sets the condition to obtain parallel paired cracks (brown line), and $\Gamma_0 = \Gamma_1$ for follower cracks (black line).
These conditions are compatible with our experiments (circles and squares represent the patterns experimentally observed) although experimental uncertainties do not allow quantitative determination of the boundaries. Crescent and paired cracks are also observed in the domain where isolated cracks are possible. 
Different states can indeed coexist for the same set of parameters depending on nucleation. 
While spirals are triggered by localized defects, crescent alleys require relatively long initial cracks to develop and parallel bands are found along rough boundaries. 
More complicated interacting structures can also be observed when the density of defects is increased (Fig~1.A). 
Nevertheless, we do not expect any fracture  when $\Gamma_0>\Gamma_1$, which defines a new, reduced domain for the stability of coatings  (in darker pink in Fig.~4). 
Systems where residual stresses increase progressively (for instance as drying takes place) follow a straight line in this diagram starting from the origin. 
The first failure mode encountered is therefore either classical isolated cracks or the collaborative modes if  $\Gamma_0/G_c<\gamma$. This condition quantifies the fact that when adhesion energy is weaker than fracture energy, delamination collaborates with transverse fracture into a cooperative failure mode.\\

\begin{figure}[!ht]
\centering
  \includegraphics[width=0.9\columnwidth]{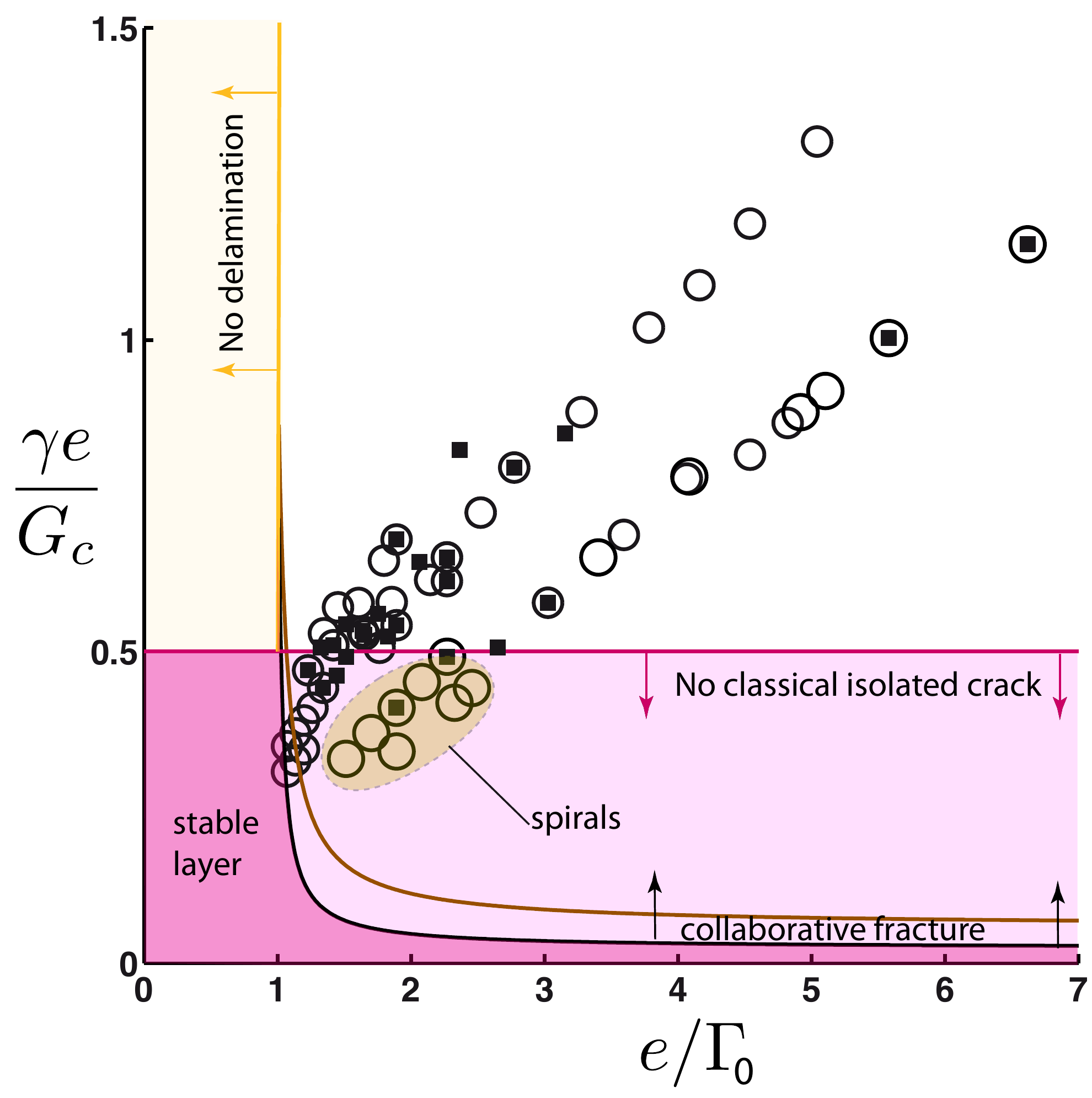}
  \vspace{-4mm}
  \caption{Experimental observations of patterns in the $(\gamma e/G_c, e/\Gamma_0)$ plane:
cracks following a previous cut ({$\bigcirc$}, 
spiral or crescent) and paired cracks leading to self peeling strip ({\tiny$\blacksquare$}).
Two borders $\gamma e/G_c \geq0.5$ (pink line) and $e/\Gamma_0 \geq 1$ (yellow line) classically define the regions where isolated channel cracks and delamination are respectively impossible.  
However, collaborative delamination cracks are observed within the light pink domain. 
The different patterns are observed for $\Gamma_2 >\Gamma_0$ (paired cracks above the brown curve), $\Gamma_1 > \Gamma_0$ (spirals and oscillating crescents above the black limit).
Different patterns are observed for the same range of physical parameters, depending on nucleation geometries. 
We observe crescent patterns with larger amplitude
 closer to the black limit. 
}
  \label{DiagConf}
\end{figure}

Although cracks are usually viewed as a failure, the extreme robustness of the  self-replicated patterns induced by the collaborative mode presented here can turn fracture into a design tool.
The patterns form spontaneously along a path determined completely by the geometry of the initiation spot
independently from inhomogeneities of adhesion properties (often difficult to control perfectly). {Recently developed micropatterning techniques allow the control of the geometry of initiation~\cite{Nam12}.} Our study provides the operating conditions for these robust patterns through a physical description of the phenomenon supported by numerical and experimental tests.

We thank ECOS C12E07, CNRS-CONICYT, and Fondecyt Grant No. N1100537 for partially funding the project. We are very grateful to M\'elanie Lebental for introducing us to such beautiful crack samples.

%

\newpage
\cleardoublepage

\noindent{\bf Supplemental Material} \\

\section{Methods: deposition and characterization of the layers}

\noindent{\bf Coating procedure} \\
Silica layers were deposited by spin-coating commercial ``spin on glass'' (SOG, Accuglass T-12B from Honeywell) methyltriethoxyorthosilicate solutions on silicon substrates (rotation speed of 500 to 1200 rpm for a duration of 18 to 22 s). 
The liquid layer was baked at $200^\circ\,$C in a oven for 2 hours, allowing the solvent to evaporate and the sol-gel condensation to operate~\cite{Lebental07}. A thin first layer of commercial SOG, methyltriethoxyorthosilicate, tetraethylorthosilicate, glycidoxypropyltrimethoxysilanetetraethoxysilane or 1H,1H,2H,2H-Perfluorodecyltrichlorosilane was deposited to modify the adhesion energy with the substrate.
The thicknesses of the layers were determined with a contact profilometer (Dektak), AFM or FEG-SEM microscope. The cracks started propagating as the samples were removed from the oven and were thus immediately observed with an optical microscope. Additional SEM or AFM observations were conducted on {\it post mortem} samples. When cracks did not nucleate spontaneously, we tried to trigger their propagation manually by scratching the sample with a blade. Although this last operation was not precisely controlled, we only expect the stress distribution to be modified in the vicinity of the scratch.\\

\begin{figure}[!h]
\centering
  \includegraphics[height=6cm]{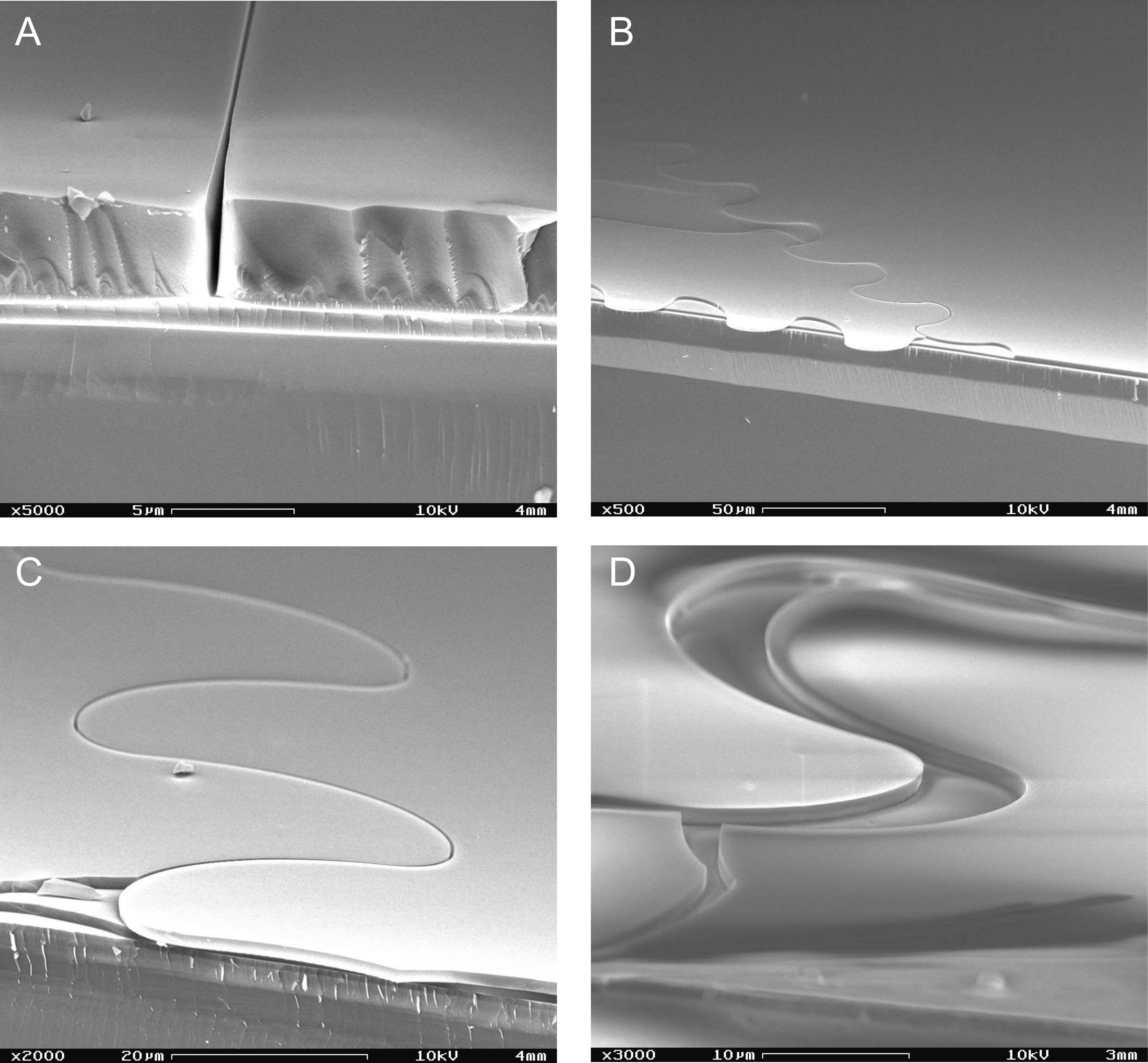}
  \caption {FEG-SEM images of the cross-section of the deposited layers. {\bf A}. Isolated channel crack in a 4$\,\mu$m thick silicate layer deposited on a silicon wafer. {\bf B-D}. Crescent alleys in a 955~nm thick silicate layer. }
 \label{FEG}
 \end{figure}

\noindent{\bf Observation by microscopy} \\
Cross sections of the layers were imaged with Field Emission Gun Scanning Electron Microscopy (FEG-SEM) for isolated channel crack and crescent alleys (Fig.~\ref{FEG}). 
The layers do not exhibit any anisotropy and the crack path is nearly orthogonal to the substrate. 
In the case of a layer deposited on a preliminary (stable) SOG coating, images in Fig.~~\ref{FEG}B-C show that only the top layer is cracking, and that delamination occurs below this first layer. \\

\begin{figure}[!h]
\centering
  \includegraphics[height=5cm]{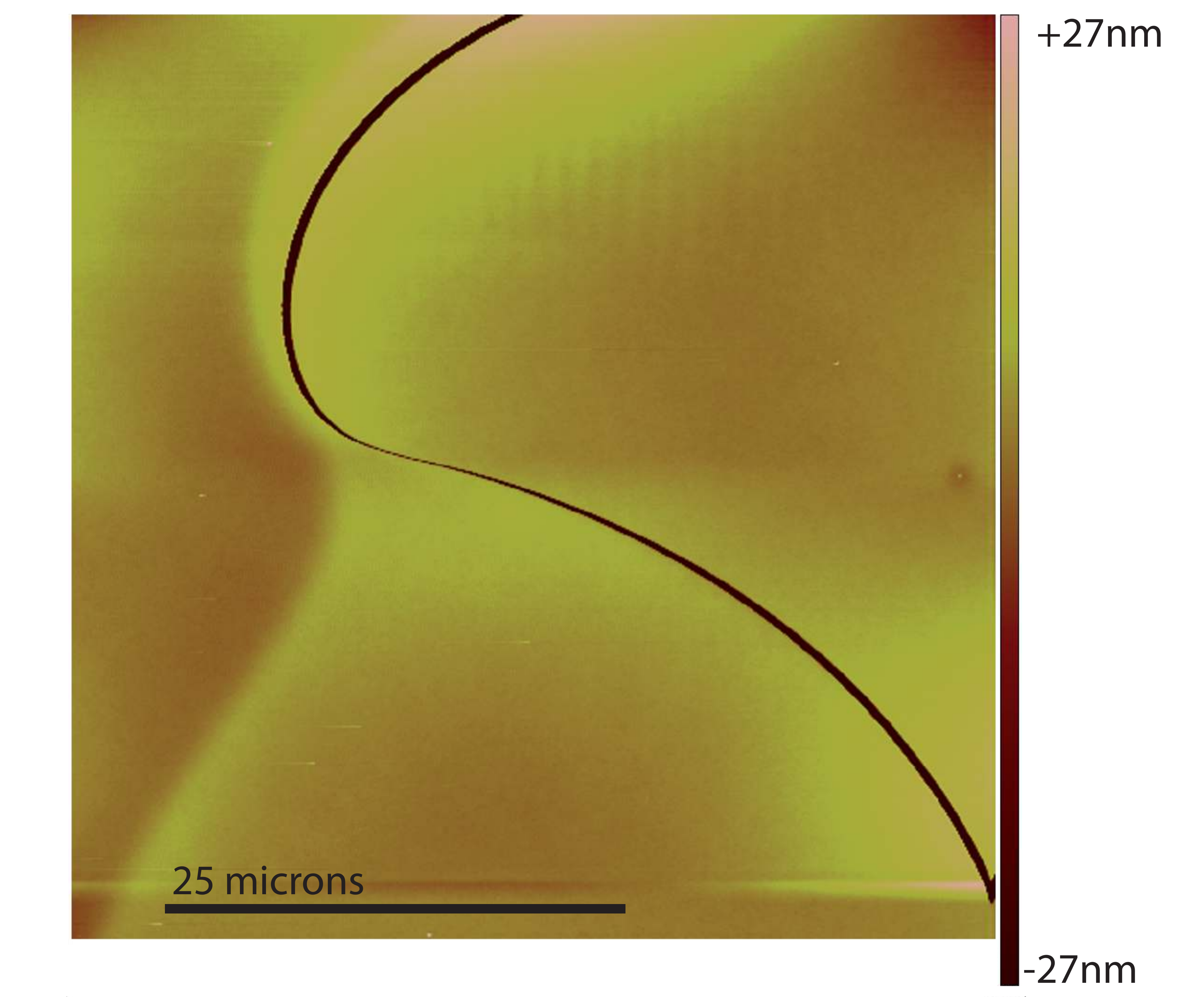}
  \caption {AFM images of a $1 \mu$m thick silicate film, revealing Poisson dilatation due to in-plane contraction.}
\label{AFM}
 \end{figure}

AFM images (Fig.~\ref{AFM}) and phase contrast microscopy reveal the variation of thickness of the layer when the stress is released in the $z$ direction. We use this information to estimate the Poisson ratio of the layer. This feature is a powerful tool to determine the area where delamination occurred even if the layer eventually re-adheres to the substrate. Note also the large opening along the crack path, resulting from contraction over large areas.   \\

\noindent{\bf Measurement of the Young modulus}\\
Once removed from the oven, the Young modulus of the film ($E_f=4 \pm 1 $GPa) was determined by nanoindentation (MTS XP form Agilent) on  layers of SOG of $2\,\mu$m of thickness spin-coated on silicon substrates.  \\

\noindent{\bf Measurement of Poisson ratio} \\
The Poisson ratio was determined by measuring the variation of the thickness of the film by AFM on a stripe delaminated from the substrate were the stress is fully released compared to a film under residual stress. 
The relative variation of thickness $\epsilon_{zz}$ is related to the Poisson ratio~$\nu$ : $\epsilon_{zz}=2\frac{\nu}{1-\nu}\epsilon_{xx}$.
For a film of $1\,\mu$m, we measure a variation of thickness of  $6\,$nm, which corresponds to a Poisson ratio of $0.25 \pm 0.05$.  \\

\noindent{\bf Measurement of the residual strain} \\
The residual strain in the film was determined by depositing the SOG coating on a thin (100) silicon wafer (of thickness of $100\,\mu$m) and subsequently measuring the slight deflection of the wafer induced by the strain. 
The curvature $\kappa$ of the wafer is related to the stress in the film through Stoney's law $\sigma=\frac{E_{s}h_{s}^2 \kappa}{6h(1-\nu_{s})}$~\cite{Stoney1909}, where $E_{s}={169}\,$GPa and $\nu_s = {0.36}$ are respectively the Young's modulus and the Poisson coefficient of the substrate. For 0.68, 0.9 and $1\,\mu$m thick films, we measured respectively curvatures of 0.1, 0.14 and 0.15 $m^{-1}$, which corresponds to a residual stress of 55 MPa and a strain on the order of $1\%$. This value is consistent with direct strain field measurement during fracture propagation.  \\

\noindent{\bf Measurement of the strain field} \\
Submicrometer scale carbon particles were sputtered on the surface of the film by placing the film a few seconds above the smoke of a candle. 
The displacement field following the motion of a crack front and the corresponding strain field were inferred from the image correlation software DAVIS from LaVision.  \\

\noindent{\bf Estimation of the fracture energy} \\
The fracture energy of the film is computed from the minimal thickness $h_c$ required for the propagation of channel cracks from initial flaws in the case of strong adhesion. 
We obtained a critical thickness $h_{c}= 1.8\,\mu$m for a residual stress $\sigma = 55\,$MPa, which leads to a fracture energy, $G_c=2 \gamma h \sigma^2 (1-\nu)/E = 1.5 \pm 0.2\,$N/m with $\gamma=0.64$ obtained numerically (the layer is here directly deposited on the rigid substrate).   \\

\begin{figure}[!h]
\centering
  \includegraphics[width=\columnwidth]{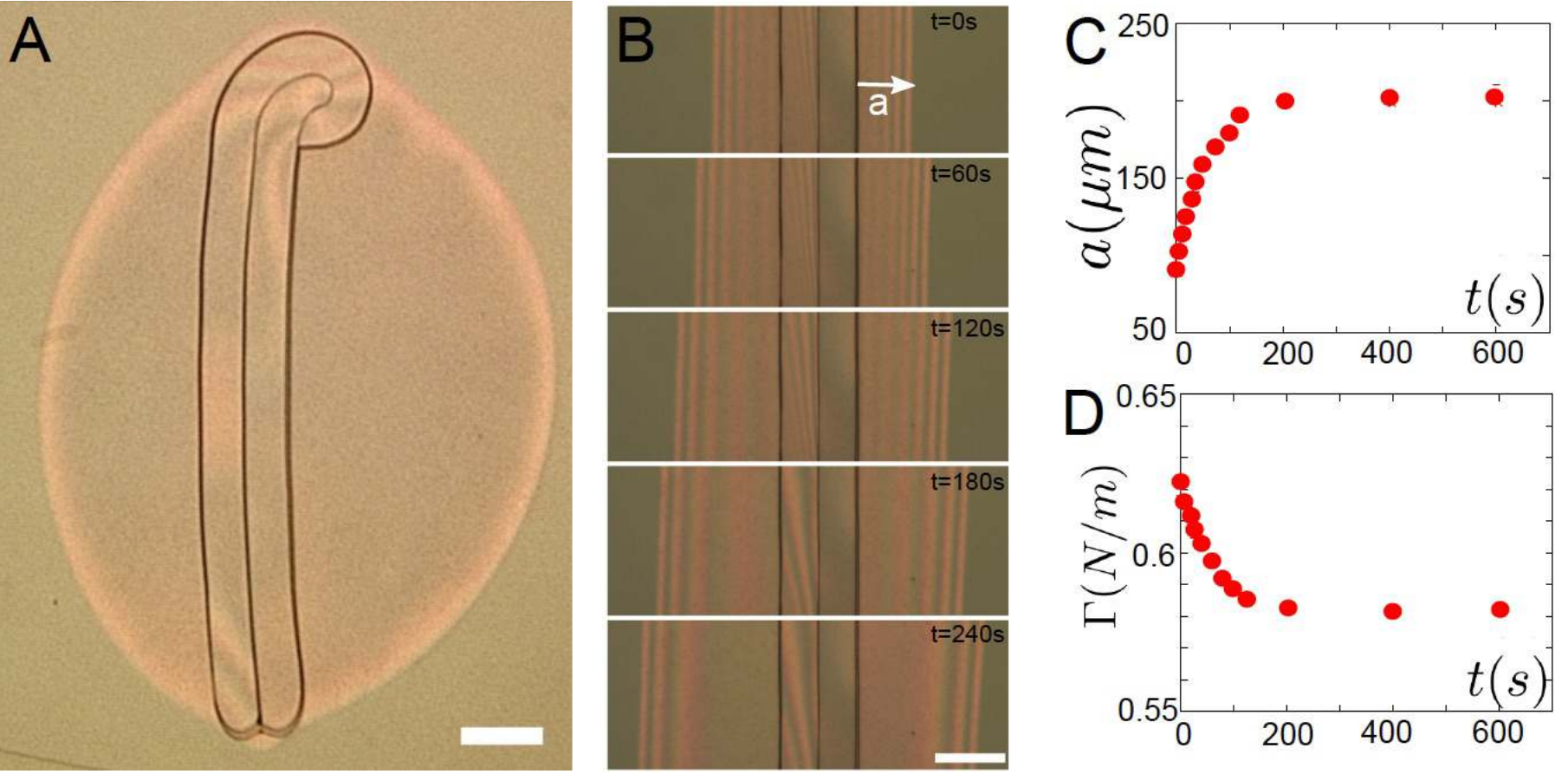}
  \caption {{\bf A-B}. Cut test experiment: the progressive delamination of the the film around a steady straight crack is monitored as a function of time (scale bar $100\,\mu m$).
{\bf C}. Evolution of the extent of the delamination front $a$ as a function of time. 
{\bf D}. The corresponding delamination energy computed from the shape of the front decreases progressively in time and reaches a plateau value $\Gamma_0$.}
  \label{fig-adhesion}
\end{figure}

\noindent{\bf Estimation of the adhesion energy} \\
The adhesion energy is deduced from the shape of the delamination front around a steady straight crack (Fig.~\ref{fig-adhesion}). 
The shape of the front depends on three parameters $\zeta$, $\lambda$ and the Poisson ratio $\nu$~\cite{Jensen90}. 
$\zeta=\frac{\sigma}{\sigma_c}$ is the ratio between the actual strain in the film $\sigma$ and the critical strain associated with steady-state plane delamination $\sigma_c$, $\lambda$ is a parameter which accounts for the influence of the contribution of mode 3. 
The fronts obtained in experiments are comparable to theoretical predictions with $\lambda=0$, suggesting that mode 3 is negligible in our case. 
Under this condition, the aspect ratio of the delaminated zone is only a function of $\zeta$ (see Fig.~8 in ~\cite{Jensen90}) which allows to estimate the adhesion energy as, $\Gamma=\frac{\sigma^2 (1-\nu^2)h}{2E \zeta}$.
The delamination front extends slowly with time (Fig.~\ref{fig-adhesion}, and Supplementary movie S5), which shows that the adhesion energy $\Gamma$ depends on time.
In other words, adhesion enegy depends on the velocity of the delamination front (Fig.~\ref{fig-adhesion}D), as it has already been observed with similar systems~\cite{Lin07}.
We refer to $\Gamma_0$ as the plateau value of the adhesion energy at long times, {\it i.e.} when the crack reaches a steady shape.

\section{Quantitative calculation using numerical simulations}

\begin{figure}
\centering
  \includegraphics[width=\columnwidth]{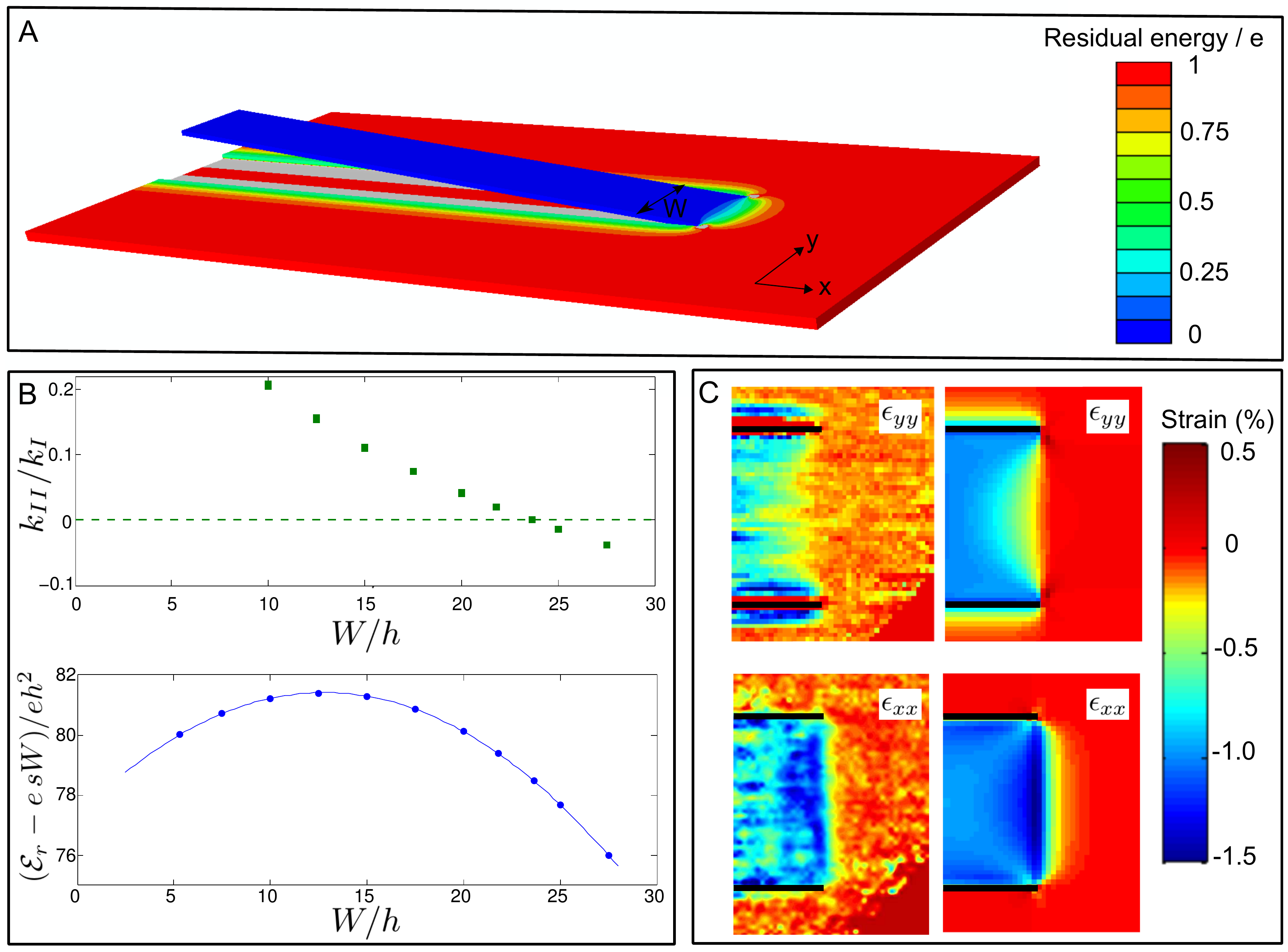}\\
  \caption{  
{\bf A,} Finite element calculations of the elastic strain energy (color coded) on a delaminated band (of optimal width $W_2$) cut from a pre-strained film adhering on a substrate. {The strip spontaneously tilts up.}
{\bf B, (top)}, {stress intensity factors ratio $k_{II}/k_{I}$ from numerical calculation} as a function of the normalized width $W/h$, where $h$ is the thickness of the coating. Since the sign of the ratio is related to the direction of propagation of the crack, parallel cracks are expected for $W \simeq 23.7h$. 
{\bf (bottom)}, Elastic energy $(\mathcal{E}_r-e\,sW)/eh^2$ computed from numerical calculation, and fit by a function $-\alpha (W/h)^2 + \beta (W/h)+c,$ {with $\alpha=0.0251,\beta=0.642.$} 
{\bf C,} Experimental measurements of the strain fields estimated {by digital image correlation} (left) and comparison with the numerical calculation (right). The component $\epsilon_{yy}$ (up) shows that the residual stress is released in the strip except  in the vicinity of the delamination front. The component  $\epsilon_{xx}$ (down) illustrat{es} the stress released {ahead of the debonding front due to the tilting of the strip}.}
\label{Fig_Numeric}
\end{figure}

Numerical simulations of  a straight debonded strip are performed within three-dimensional linear elasticity (Abaqus software, Dassault Syst\`emes).
Guided by experimental observation, we impose a straight debonding front (Fig.~\ref{Fig_Numeric}A).
 The delamination front is straight, perpendicular to the cracks separated by a width $W$. The numerics includes the first layer covering an infinitely stiff  substrate.
The computation is validated by comparing computed and experimental estimations of the strain fields (Fig.~\ref{Fig_Numeric}C).
 
The computed elastic energy of this system $\mathcal{E}_r$ for different widths $W$ provides the energy released during propagation. The difference 
 $(\mathcal{E}_r-e\,sW)/eh^2$ is well fitted by a function $\alpha (W/h)^2 - \beta (W/h)$, with $\alpha \simeq 0.0251 $ and $\beta \simeq 0.642$.
 
Numerical simulations also provide a direct computation of stress intensity factors at the crack tips, as a function of $W$, from which the direction of the crack can be predicted. 
As stated in the principle of  local symmetry, a crack is expected to follow a straight path if the shear stress intensity factor $k_{II}$ at its  tip vanishes~\cite{Hutchinson92}. 
In the numerics, this condition is only satisfied for a given value $W_2=23.7h$ (Fig.~\ref{Fig_Numeric}B), in quantitative agreement with  our experiments in SOG coatings where $W_2\sim 25 h$. 
This path is stable since the sign of $k_{II}$ indicates an inward (respectively outward) propagation for $W > W_2$ (resp. $W<W_2$).\\

However, we note that the values of $\alpha$ and $\beta$ lead to an optimal width $W_2/h= 12.78$ in Eq.(3), 
which is smaller than prediction from stress fields, and than observed in experiments. 
The discrepancy in the theoretical results  is due to an approximation in the computation of the elastic energy released along the crack path. We have indeed assumed 
a released elastic energy proportional to  the length of the path $\gamma h s$, independently from the curvature of the path.
The value of prefactor $\gamma$ was actually obtained by numerical computation of a straight cut where the stress field is invariant along the crack line (plane strain).
This is not true if the path is curved, leading to a three-dimensional stress field.
We give here an estimate of this effect, and show  that it tends to increase the equilibrium width $W_2$ by a significant amount,
reconciling both theoretical approaches.

\begin{figure}[h]
\centering
  \includegraphics[width=\columnwidth]{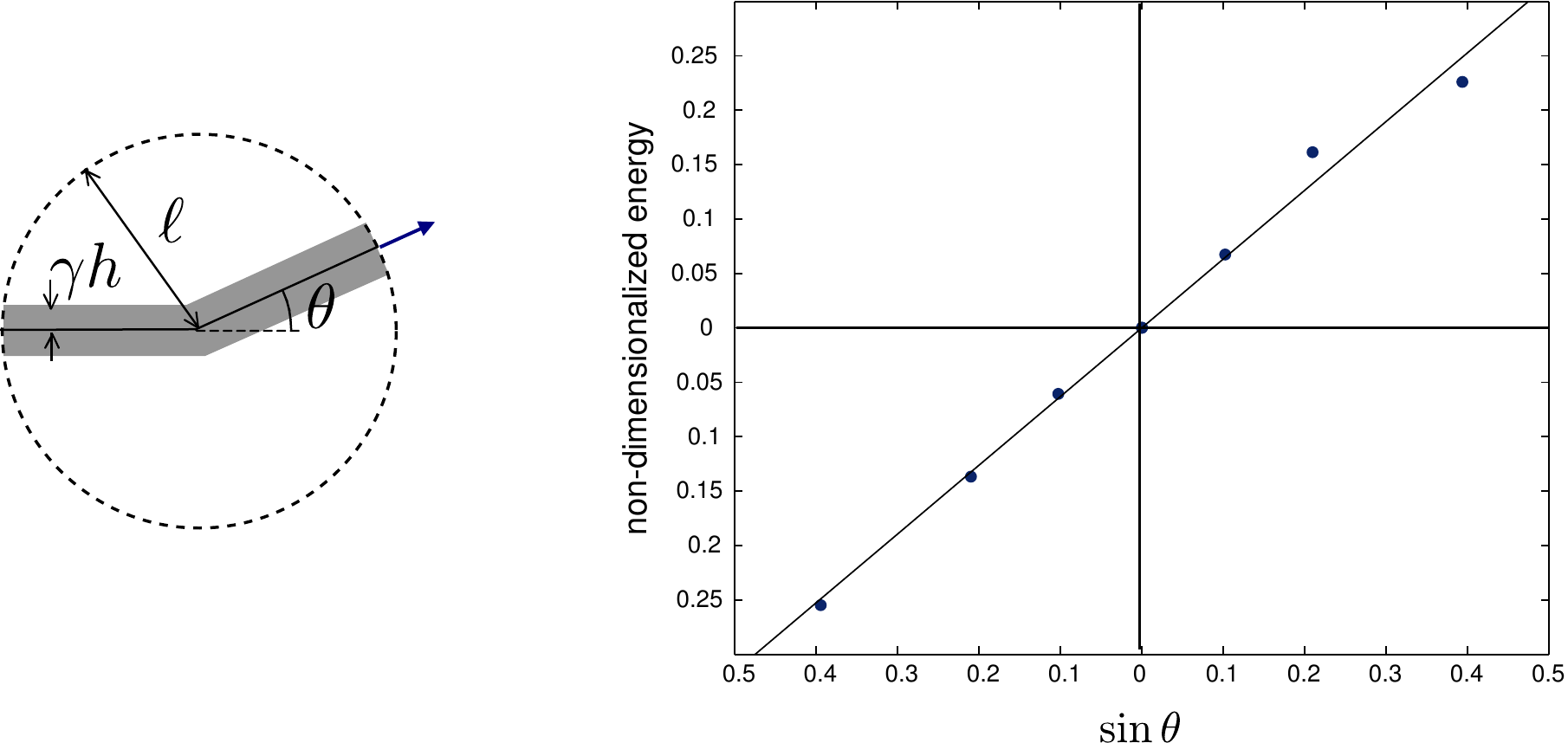}\\
  \caption{{\bf A}. Kink of angle $\theta$ in the crack path. {\bf B}. Additional energy released due the kind non-dimensionalized by $eh^2$}
\label{fig_beuth}
\end{figure}

We now compute the perturbation to the elastic energy released due to a kink of the crack path with an angle $\theta$. 
If we locate the kink at the center of a disk of radius $l\gg h$, the crack path separates the disk into two sectors. 
According to our approximation, the elastic energy released in each sector, $\mathcal{E}_{rs}(\theta)= \gamma h l$, is independant of $\theta$. 
However, numerical computations show that more energy is released in the left sector when $\theta$ is positive (and less in the opposite situation).  
This effect was not included in our description, and favors the outward propagation of the strip. 
Nevertheless, we note that in the case of classical isolated channel cracks, adding up the contribution of left and right leads to a maximum energy release rate for $\theta=0$. As a consequence, isolated channel cracks are not expected to form kinks, but to propagate along straight paths, as observed in experiments.

A reasonable fit for this numerical result (valid for $\theta \ll 1$ and $l\gg h$) is $\mathcal{E}_{rs}(l,\theta) = e( \gamma h l + \delta h^2 \sin\theta )$, where $\delta$ is of order 0.63. 
It is not straightforward to include this effect in Griffith's relation, which considers an infinitesimal propagation of a kinked crack and includes the corresponding energy release rate $\partial \mathcal{E}_{rs}(l,\theta) /\partial l.$

This quantity is not directly measurable because of the necessary condition $l\gg h$ in our numerical approach. 
However,  we assume as a first approximation that this energy correction is released when the crack propagates by a distance on the order of a thickness $h$, leading to a term $\delta h \sin\theta dl $, with an unknown prefactor of order one.
Griffith's relation (Eq.~3) is thus modified into:
$$
d{E}/dl = e [ W\cos\theta + 2\gamma h + 2h \delta  \sin\theta -2(2\alpha W - \beta h)\sin\theta ] 
$$
We note that this additional term is equivalent to change $\beta$ into  $\beta +\delta$.
Since  this correction is difficult to compute exactly, we take $\beta=1.26$ which corresponds to the width $W_2/h\sim 23.7$ obtained in the numerical study of the stress intensity factor. We note that the correction $ \delta = 0.56$  is compatible with the order of magnitude estimate $\delta \sim 0.63$

As a conclusion, our simple energetic approach is consistent with the direct numerical calculation of stress intensity factors, using a value of parameter $(\alpha=0.0251,\beta=1.26)$ which we use throughout the article. 

\section{Crack following a previous one : spirals and crescents}
\begin{figure}[h]
\centering
  \includegraphics[width=6cm]{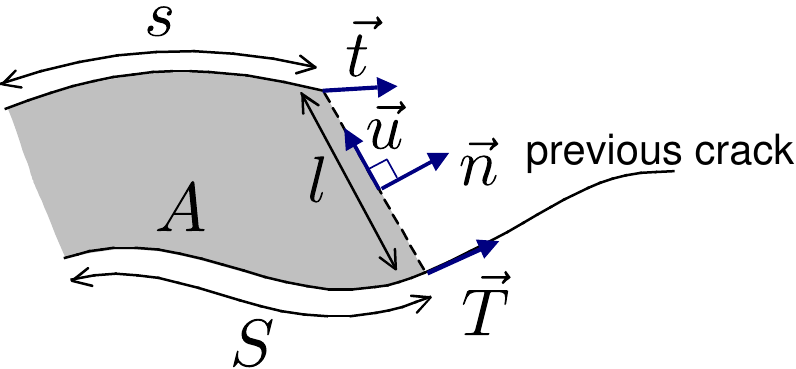}\\
  \caption{Front following a previous crack at a distance $l$. The old and new cracks are characterized by paths $S$ and $s$ and local tangents $\vec T$ and $\vec t$, respectively. $\vec u$ and $\vec n$ define the tangent and normal directions to the delamination front of width $l$.}
\label{Fig:following}
\end{figure}

In spiral and crescent morphologies, an advancing fracture follows an older crack path. The debonding front connecting the crack tip of abscissa $s$ to a point of abscissa $S$ along previous arbitrary cut has now to be determined (Fig.~\ref{Fig:following}). We assume here that the crack front is a straight segment with length $l$, normal and tangent vectors $\vec n$ and $\vec u$, respectively.\\

The elastic energy released is now $\mathcal{E}_r=eA + \gamma ehs + ef(l) $, where $f$ accounts for the contribution of the region of the film neighboring the delamination front.  We assume that $f$ takes the same functional form as in the previous case, $f(l)= -\alpha l^2 + \beta hl$. Geometry now implies variations of the area $ 2dA = l\vec n \cdot (\vec t ds + \vec T dS)$ and of the front length $ dl = \vec u \cdot (\vec t ds - \vec T dS)$, where $\vec T$ and $\vec t$ correspond to the local tangent to the old and new paths, respectively. 
Griffith's criterion reads   $d\mathcal{E}_r=\Gamma dA + G_c h ds$, and leads to: 
$$
(C\vec n  -D\vec u)\cdot \vec T\,dS  +  \left[(C\vec n + D\vec u)\cdot \vec t - (G_c - \gamma e)h\right]ds=0 ,
$$
with
$ C =  (e- \Gamma)l/2 $
and
$ D = ef'(l) = \beta e h -2\alpha e l .$
Since $S$ and $s$ are independent variables, both terms {must} vanish.
The first equation,
\begin{equation}
(C\vec n - D\vec u) \cdot \vec T =0
 \label{eq:1crackfront}
\end{equation}
 imposes the geometry of the debonding front ({\it i.e.} the position of point $S$). 
The second equation, 
\begin{equation}
\gamma  e h + (C\vec n  + D\vec u  ) \cdot \vec t  = G_c h,
 \label{eq:1crackGr}
\end{equation}
is equivalent to Griffith's relation.
According to the maximum energy release rate criterion, fracture propagates in direction: 
\begin{equation}
 \vec t \mbox{ parallel to } ( C\vec n + D\vec u  ).
 \label{eq:1crackdir}
\end{equation}

The set of rules (\ref{eq:1crackfront}-\ref{eq:1crackdir}) allows to compute the trajectory of a crack when propagating together with a delamination front bounded by a previous cut of arbitrary geometry. Fig.~3 compares several qualitatively different experimental paths with predictions using the parameters $(\alpha,\beta,\gamma)$ determined previously,  starting from different initiation geometries:  an Archimedian spiral (Fig.~3B) starts from a localized defect; an alley of crescents (Fig.~3D) {is generated from an initial cut extracted from experimental path;}
 three interacting crack paths (Fig.~3C) {are computed from three segments intersecting at 120$^o$ degrees}. The {calculated} fracture paths reproduce the diversity of geometry of experimental fracture trajectories with a remarkable precision given the simplicity of the propagation rules.

In all situations, following fractures tend to propagate at a well defined distance from the partner cut. This distance can be analytically assessed in two simple pre-existing cut geometries: a straight line and a point. Parallel propagation along a straight partner cut implies $\vec t = \vec T$, which leads to $C=-D= (G_c-\gamma e)h/\sqrt{2},$ and to a $45^{\circ}$ tilt angle of the delamination front. 
The width of the delaminated strip $W_1$, and propagation velocity $v$ are finally given by $\Gamma(v)=\Gamma_1$ where:
\begin{eqnarray}
W_1 &=& \frac{\beta h}{2\alpha}\left(\frac{1}{\sqrt{2}}+\frac{G_c/e  - \gamma} {2\beta} \right) \label{eq:W1}\\
 \frac{\Gamma_1}{e} &=&  1 -  \left(\frac{G_c}{e}-\gamma \right)\frac{h}{W_1} \label{eq:speedW1}
\end{eqnarray}
Eq.~(\ref{eq:W1}) gives the right order of magnitude for $W_1$, however the  quantitative dependence of $W_1/h$ with $G_c/e$ is not directly evidenced in our experiments (Fig.~2A). 

%

\end{document}